\documentstyle[12pt]{article}
\newcommand{\be}{\nopagebreak\begin{equation}}
\newcommand{\ee}{\end{equation}}
\newcommand{\ba}{\begin{array}}
\newcommand{\ea}{\end{array}}
\newcommand{\bp}{\begin{picture}}
\newcommand{\ep}{\end{picture}}
\newcommand{\eol}{\\[1pc]}
\newcommand{\rf}[1]{Ref.~\cite{#1}}
\newcommand{\eq}[1]{Eq.~(\ref{#1})}
\newcommand{\bi}[6]{\bibitem{#1}{#2, }{\sl #3 }{\bf #4}{ (#5)}{ #6}}
\newcommand{\wt}[1]{\widetilde{#1}}
\newcommand{\wh}[1]{\widehat{#1}}
\newcommand{\ie}{{\em i.e., }}
\newcommand{\eg}{{\em e.g., }}
\renewcommand{\d}{\partial}
\newcommand{\eps}{\varepsilon}
\newcommand{\scr}{\scriptstyle}
\newcommand{\scs}{\scriptscriptstyle}
\newcommand{\R}{{\scriptscriptstyle R}}

\renewcommand{\thefootnote}{\fnsymbol{footnote}}

\begin{document}
\begin{titlepage}
\begin{flushright}
NBI-HE-93-67\\
November, 1993
\end{flushright}
\vspace*{36pt}
\begin{center}
{\huge \bf 3D GRAVITY AND GAUGE \\[1pc]
THEORIES\footnote[1]{Contribution to Proceedings
of 1993 Cargese workshop}}
\end{center}
\vspace{2pc}
\begin{center}
 {\Large D.V. Boulatov}\\
\vspace{1pc}
{\em The Niels Bohr Institute\\
University of Copenhagen\\
Blegdamsvej 17, 2100 Copenhagen \O \\
DENMARK}
\vspace{2pc}
\end{center}
\begin{center}
{\large\bf Abstract}
\end{center}
I argue that the complete partition function of 3D quantum
gravity is given by a path integral over gauge-inequivalent manifolds
times the Chern-Simons partition function. In a discrete version, it
gives a sum over simplicial complexes weighted with
the Turaev-Viro invariant.
Then, I discuss how this invariant can be included
in the general framework of lattice gauge theory (qQCD$_3$). To make
sense of it, one needs a quantum analog of the Peter-Weyl theorem
and an invariant measure, which are introduced explicitly. The consideration
here is limited to the simplest and most interesting case of $SL_q(2)$,
$q=e^{i\frac{2\pi}{k+2}}$. At the end, I dwell on 3D generalizations of matrix
models.
\vfill
\end{titlepage}

\section{Introduction}

During the last few years, considerable progress has been made in our
understanding of 2D quantum gravity and string theory (see review
\cite{David} and references therein). What helped greatly to fight
the problem was the fortunate interplay of the methods of conformal
field theory and the computational power of matrix models. For those who
tries to think of quantum gravity seriously the next step has naturally
been the path integral over 3D manifolds. It is not a priori
doomed-to-fail enterprise. Indeed, although the problem is really a hard one,
some interesting results have already been obtained. As well as in the
2D case, there are essentially two approaches. The first starts with a
continuous formulation trying to make sense of a path integral over
metrics. The main achievement on this way has been the connection with the
Chern-Simons theory established by E.Witten \cite{Witten}.
The second approach is based completely on lattice
experience. Here the path integral is substituted by a sum over all
simplicial (or another kind of) complexes. The gained advantage
is the finiteness of all involved quantities and relative
simplicity, which allows for numerical investigations.
However, the main problem, native to all lattice models, is what kind
of continuum limit (if any) can be reached in every particular case?

In the present paper, I try to establish a connection between these two
approaches, paying more attention to the second one, however.

In Section~2 I remind a reader the basic notions of 3D general relativity and
describe its connection with the Chern-Simons theory.

Section~3 is devoted to 3D simplicial gravity. I formulate the model and
review some results of numerical investigations.

In Section~4 I define qQCD$_3$ and show that its weak-coupling limit is
related to the Turaev-Viro invariant.

In Section~5 a model which can be regarded as a 3D generalization of
the one-matrix model are introduced.

Section~6 contains some general remarks.

\renewcommand{\thefootnote}{\alph{footnote}}

\section{3D gravity and Chern-Simons interpretation}

The partition function in Euclidian quantum gravity is intuitively
defined as a sum (path integral) over all manifolds weighted with the
exponential of a reparametrization invariant action

\be
{\cal P}=\sum_{{\cal M}^3} e^{S}
\label{parfun}
\ee

By definition, the manifold is a topological space which can be globally
covered with local coordinate systems. In other words, every its point
has an open vicinity allowing for a continuous one-to-one map into
$R^3$. On a manifold, one can define functions, vector fields, forms and
tensors. To make sense of the partition function (\ref{parfun}),
a metric tensor,
a volume form and an affine connection are needed.
The metric tensor is a scalar
bi-linear symmetric function on vectors, \ie a second-rank contra-variant
tensor $g_{ij}$. The matrix $g_{ij}$ has to be invertible:
$g_{ij}g^{jk}=\delta^k_i$. If $g_{ij}\equiv 0$ on some sub-manifold, it
should be regarded as non-compactness. Without metric one cannot define the
functional integral measure.

The volume form is some fixed
3-form\footnote{In what follows, for convenience,
I denote forms with the tilde and
vector fields with the hat, \eg $\wh{\d}_i\equiv \frac{\d}{\d x^i}.$}, $\wt V$.
It is always convenient to make it compatible with the metric. Then,
in coordinates,

\be
\wt V = \sqrt g dx^1\wedge dx^2\wedge dx^3
\ee
To choose a coordinate basis means to fix 3 mutually commutative vector
fields $\wh{\d}_1,\ \wh{\d}_2,\ \wh{\d}_3:\ [\wh{\d}_i,\wh{\d}_j]=0$.
Sometimes, it is convenient to have a non-coordinate basis $\{\wh e_a\}$:

\be
[\wh e_a, \wh e_b]=C^c_{ab} \wh e_c
\label{[e,e]=}
\ee

Let me choose it such that

\be
g_{ij}e_a^ie_b^j = \delta_{ab}
\ee
where $e_a^i$ are the components: $\wh e_a=e_a^i\wh{\d_i}$. I shall
refer to them as the dreibein.

To introduce the Riemann tensor one needs the notion of an affine
connection, which defines rules of a parallel transport of vectors:
$\nabla_{\wh e_a} \wh e_b = \omega^c_{ba} \wh e_c$.

If one uses
forms\footnote{The groups of indices $abc$ and $ijk$ belong to
different bases!},
$\wt e^a=e^a_i dx^i$ and $\wt{\omega}^a_b=\omega^a_{bi}dx^i$,
one can introduce the Riemann tensor
$\wt R^a_b=\frac12 R^a_{b,ij} dx^i\wedge dx^j$
and the torsion $\wt T^a=\frac12 T^a_{ij} dx^i\wedge dx^j$ in the most
elegant way (Cartan's structural equations)

\be
\wt R^a_b = {\rm d}\wt{\omega}^a_b - \wt{\omega}^a_c\wedge\wt{\omega}^c_b
\ee

\be
\wt T^a = {\rm d}\wt e^a -\wt{\omega}^a_b\wedge \wt e^b
\ee
If the torsion vanishes, the connection is said to be symmetric.
In this case, it is determined by the
commutator (\ref{[e,e]=})

\be
\omega^c_{ab} = \frac12(C^a_{cb}+C^b_{ca}-C^c_{ab})
\label{omega=}
\ee

The Einstein-Hilbert action can be written in the
form

\be
S=\lambda \int \epsilon_{abc}\wt R^a_b\wedge \wt e^c +\beta\int
\wt e^1\wedge \wt e^2\wedge \wt e^3
\label{RHaction}
\ee

Witten suggested to consider the dreibein and the
Levi-Civita connection as the gauge variable:

\be
A_i = e^a_iP_a + \omega ^a_{bi}J_{ab}
\label{gf}
\ee
taking values in the $ISO(3)$
Lie algebra (if the signature is Euclidian and the cosmological constant
is zero):

\be\ba{l}
{[J_{ab},J_{cd}]=\delta_{ac}J_{bd}+\delta_{bd}J_{ac}
-\delta_{bc}J_{ad}-\delta_{ad}J_{bc}} \eol
{[J_{ab},P_c]=P_a\delta_{bc}-P_b\delta_{ac}} \eol
{[P_a,P_b]=0}
\ea
\ee
with the invariant metric on the algebra: $\langle P_a,P_b\rangle =
\langle J_{ab}, J_{cd}\rangle = 0$, $\langle P_a,J_{bc}\rangle =
\epsilon_{abc}$.

The obvious problem here is: what meaning do we give to the generators?
If $P_a$'s are to represent vector fields, $\wh P_a$, forming a
coordinate basis (they commute), then $[e^a_i\wh P_a,e^b_j\wh P_b]$ is
not zero except for the case when the space is flat and the dreibein
appears as a coordinate transformation: $e^a_i=\frac{\d y^a}{\d x^i}$.
In a curved space, the Lie algebra generators have an indefinite
meaning.

However, the construction is not so restrictive as might seem from
this consideration.

Let $\wh v=v^i\wh{\d}_i=\gamma^a\wh e_a$ be an infinitesimal vector
field.
The variation of the basis $\wh{\d}_i$ under the diffeomorphism
generated by $\wh v$ is given by the Lie derivative

\be
\pounds_{\wh v}\wh{\d}_i\equiv [\wh v,\wh{\d}_i] = \nabla_{\wh v}\wh{\d}_i
- \nabla_{\wh{\d}_i}\wh v = (-\nabla_i\gamma^a + \omega^a_{bj}v^j e^b_i +
v^je^a_{i,j})\wh e_a
\ee
where the comma means the derivative with respect to $x^j$.
The first equality holds if the torsion tensor identically vanishes.
So, from the view-point of the fixed non-coordinate basis $\wh e_a$, the
variation of the basis $\wh{\d}_i$ consists of (i) a ``gauge
transformation'' $\nabla_i\gamma^a$, (ii) a ``Lorentz
rotation'' $\omega^a_{bj}v^j e^b_i=\tau^a_be^b_i$ and (iii) a coordinate
shift $v^je^a_{i,j}\approx e^a_i(x+v)-e^a_i(x)$. The last term
can be removed by ``pulling back'' $e^a_i$ to its initial point in the
$x$-frame as a scalar function.

\index{Diffeomorphism}
Of course, it just repeats the famous Witten's argument \cite{Witten} that
diffeomorphisms generated by vector fields
can be regarded on-shell as gauge transformations of the
field (\ref{gf}). Maybe, it should be stressed here that one is
restricted to {\em reparametrizations}, which are not the most general
transformations. In particular,  they do not affect the commutators
(\ref{[e,e]=}).

The complete algebra of vector fields is infinite
dimensional, since the structure constants $C^a_{bc}$ are arbitrary
functions of coordinates. Hence, it cannot as a whole be reduced to any
finite dimensional symmetry, if one insists on the interpretation
of its generators as vector fields.
However, we have seen that reparametrizations can
be regarded as the gauge transformations. One can, in principle, get rid
of them by fixing a gauge and pulling out of the path integral a volume
they produce. For compact manifolds, this volume gives a topological
invariant (up to some trivial (but maybe infinite) factor). Indeed, one
can choose an arbitrary background metric, and the most convenient choice is
a solution to the Einstein equation (classical vacuum). In this case, one finds
an integral over flat connections. Witten
has noticed that, if one considers the dreibein and connection as
independent variables, the Riemann-Hilbert action (\ref{RHaction}) takes
the form of the Chern-Simons one for the group $SO(4)$. In this case,
the vanishing torsion and the Einstein equation are implied by the equations
of motion and one finds that the gauge volume should be given by the
Chern-Simons partition function. \index{Chern-Simons}
Off-shell, of course, any equivalence between
diffeomorphisms and the gauge transformations disappears.

The non-renormalizability of 3$D$ gravity means that one should work
within a regularization scheme, the choice of which can be crucial (\ie
answers will vary from scheme to scheme drastically).

\section{Quantum Regge calculus}
\index{Regge calculus}

The heuristic consideration of the previous section serves to support
the following substitution for the path integral over all 3$D$
geometries:

\be
P=\sum_{\rm topologies} I_{TV} \sum_C e^S
\label{defP}
\ee
where the first sum goes over all topologies; $\sum_C$ is the sum over
all simplicial complexes of a given topology; $I_{TV}$ is the
Turaev-Viro invariant \cite{TV};
$S$ is a lattice action, which can be taken in
the form

\be
S=\alpha N_1 - \beta N_3
\label{Slatt}
\ee
($N_k$ is the number of $k$-dimensional simplexes in a complex).

The Turaev-Viro invariant is the most reasonable substitution for the
gauge volume. For a negative cosmological constant, one finds $SO(4)$
Chern-Simons theory and, as $SO(4)=SU(2)\times SU(2)$, $I_{TV}$ seems
to be the most appropriate candidate \cite{Ooguri}.
Its ``classical'' limit was
investigated long ago by Ponzano and Regge \cite{PR}
in the framework of the Regge calculus \cite{Regge}.
Provided a triangulation is fixed, it describes an integral
over lengths of all links with a weight equal to
an exponential of the discretized Einstein-Hilbert action.
The Turaev-Viro invariant in this context may be regarded simply as
a regularization of the Ponzano-Regge construction.

As all lengths are included in $I_{TV}$, we can choose every tetrahedron
in $\sum_C$ to be equilateral. This sum serves as a natural
regularization of the path integral over classes of gauge-inequivalent
manifolds.

Using reparametrizations, one can make lengths of the
dreibein vectors equal to unity, three remaining local degrees of
freedom being angles between them.
As usual, on a lattice, one should work with a group rather than an
algebra. It means that, instead of dreibeins, their integral curves have
to be considered. In a discrete version, one fixes a finite number of
the curves going from every vertex and associate them with links of a
lattice. It is convenient to make lengths of all links equal to one
another. In simplicial complexes, angles between them
are quantized, which leads to a quantized total curvature. The Regge calculus
gives the expression for it

\be
\int d^3x\; \sqrt g R = a\Big(2\pi N_1 -6N_3\arccos \frac13\Big)
\ee
$a$ is a lattice spacing.

For manifolds, the Euler character vanishes $\chi = N_0-N_1+N_2-N_3=0$.
Together with the constraint $N_2=2N_3$, it implies that a natural
action (linear in the numbers of simplexes) depends on two free
parameters which should be related to bare cosmological and Newton
constants.

To simulate all geometries, one has to sum over all possible complexes.
Indeed, if a triangulation is fixed, commutators of lattice shifts
(analogs of the structure constants in \eq{[e,e]=}) are fixed. {\em
Fluctuating geometry assumes a fluctuating lattice}.

If a topology is fixed, the sum over simplicial complexes can be
investigated numerically. Any two complexes of the same topology can be
connected by a sequence of moves shown in Figure~1. The first move is
called the triangle-link exchange: the common triangle of two tetrahedra on
the left of Figure~1(a)
is removed and three new triangles sharing the new link appear
on the right. It increases (the inverse one decreases) the number of
tetrahedra by 1. The second move consists in the subdivision of a
tetrahedron: 4 new tetrahedra fill an old one. The inverse move is
seldom possible. However, to perform it,
one can always decrease the coordination number
of a vertex by applying the triangle-link exchange.
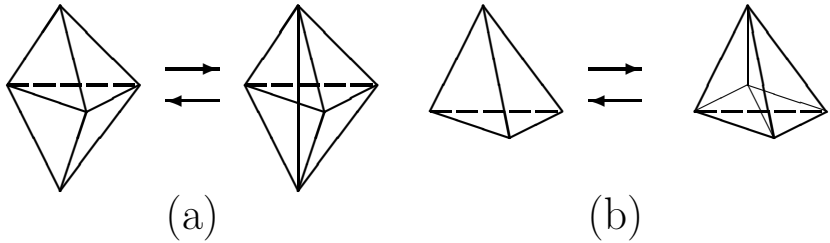
\begin{figure}
\setlength{\unitlength}{2pt}
\begin{picture}(170,100)

\thicklines
\put(10,40){\line(2,3){10}}
\put(10,40){\line(3,-1){15}}
\put(10,40){\line(1,-2){10}}
\put(20,55){\line(1,-4){5}}
\put(20,55){\line(1,-1){15}}
\put(20,20){\line(1,3){5}}
\put(20,20){\line(3,4){15}}
\put(25,35){\line(2,1){10}}
\multiput(10,40)(5,0){5}{\line(1,0){4}}

\put(40,43){\vector(1,0){10}}
\put(50,37){\vector(-1,0){10}}

\put(55,40){\line(2,3){10}}
\put(55,40){\line(3,-1){15}}
\put(55,40){\line(1,-2){10}}
\put(65,55){\line(1,-4){5}}
\put(65,55){\line(1,-1){15}}
\put(65,20){\line(1,3){5}}
\put(65,20){\line(3,4){15}}
\put(70,35){\line(2,1){10}}
\multiput(55,40)(5,0){5}{\line(1,0){4}}
\thinlines
\put(65,20){\line(0,1){35}}

\put(40,10){\makebox(10,10){\Large (a)}}

\thicklines
\put(90,35){\line(1,2){10}}
\put(90,35){\line(3,-1){15}}
\put(100,55){\line(1,-5){5}}
\put(100,55){\line(3,-4){15}}
\put(105,30){\line(2,1){10}}
\multiput(90,35)(5,0){5}{\line(1,0){4}}

\put(120,43){\vector(1,0){10}}
\put(130,37){\vector(-1,0){10}}

\put(140,35){\line(1,2){10}}
\put(140,35){\line(3,-1){15}}
\put(150,55){\line(1,-5){5}}
\put(150,55){\line(3,-4){15}}
\put(155,30){\line(2,1){10}}
\multiput(140,35)(5,0){5}{\line(1,0){4}}
\thinlines
\put(150,40){\line(0,1){15}}
\put(150,40){\line(3,-1){15}}
\put(150,40){\line(1,-2){5}}
\put(150,40){\line(-2,-1){10}}
\put(120,10){\makebox(10,10){\Large (b)}}
\end{picture}
\caption{The triangle-link exchange (a) and the subdivision (b)}
\end{figure}

Monte-Carlo
simulations using these moves as basic ``infinitesimal steps'' appear to
be quite efficient.

I do not intend to give a review of the numerical results here. An
interested reader is referred to the original papers
\cite{3dsim,3dsim2,3dsim3}.
However, a few words should be said. All simulations so far have
been carried out for the spherical topology of complexes.
It appears that the number of spherical complexes of
a given volume, $N_3$, is exponentially bounded as a function of $N_3$
for an arbitrary value of $\alpha$. It means that the definition
(\ref{defP}) is reasonable and $P$ hopefully has an appropriate
continuum limit.
\index{Monte-Carlo simulations}

One of the most
interesting observations is the resolution of the problem of the
unboundedness of the Riemann-Hilbert action within the discrete model.
As \eq{Slatt} is linear in $N_1$, it seems that most probable
configurations should be those having the maximum mean curvature, but
they are surely lattice artifacts.
However, it happened that, at $N_3$ and $\alpha$ fixed,
the probability distribution
for $N_1$ has roughly speaking the gaussian shape

\be
P_{N_3,\alpha}(N_1)\approx e^{-\frac{(N_1-\langle N_1\rangle)^2}{2\sigma^2}}
\ee
It means that, varying $\alpha$, one just shifts a position of the
maximum. Moreover, in Refs.~\cite{3dsim2}, a first order phase
transition was found at some critical value, $\alpha_c$. In the ``hot''
phase ($\alpha<\alpha_c$), crumpled manifolds dominate the partition
function. This phase is clearly unphysical. In the ``cold'' phase
($\alpha>\alpha_c$), it happens that

\be
\langle N_1\rangle (N_3)=c_1N_3+c_2
\ee
is a linear function; $c_1$ is a constant smoothly depending on
$\alpha$. Hence, the mean curvature per unit volume makes sense in the large
volume limit. However, after the naive rescaling, one finds that its
value tends to the infinity in the continuum limit. But, the
total curvature can not be regarded as an observable in quantum gravity. In
the Einstein-Hilbert action, there are two terms with dimensionful
coupling constants in fronts of them. After a regularization, one
finds the action (\ref{Slatt}) (or similar) where the total curvature has
lost its individuality and is mixed with the volume. Let us imagine a kind
of renormalization group procedure: one increases a cut-off and
integrate over fluctuations inside the blocks. The additive nature of
the total curvature means that it should undergo an additive
renormalization as well as a multiplicative one. It is natural to kill
the first by shifting the cosmological constant. Therefore, the
mean value of the total curvature is scheme-dependent and only
fluctuations make sense.

Four dimensional numerical simulations show a similar picture
\cite{4dsim}. The phase
transition there is, presumably, of the second order which might be an
evidence for graviton-like (\ie long-range) excitations in the system.

Here, the following comment is in order. One can easily obtain a
continuous manifold from a simplicial complex by using a piece-wise
linear approximation and then to smooth it. In three dimensions, any
continuous manifold allows for a unique differentiable structure and
{\em vice versa}. It means that the continuous and discrete models are
hopefully equivalent. In four dimensions, the situation is much more
complicated \cite{Nash} and it is unclear whether an entropy of smooth
manifolds can be correctly estimated within a lattice approximation.
However, simplicial gravity is interesting in its own rights. One can
simply say that, at the quantum level, the notion of the continuous
manifold is more fundamental than of the smooth one.

\section{q-deformed lattice gauge theory (qQCD$_3$)}

\index{Gauge theory! q-deformed}

In this section, I would like to show that the Turaev-Viro invariant can
be interpreted as a lattice gauge model (although a not quite standard
one). Let me start with reminding basic facts about lattice
QCD \cite{Wilson}.

Given a $d$-dimensional lattice,
a gauge variable $g_{\ell}$ taking values in a compact group $G$
is attached to each 1-dimensional link, $\ell$,
and the Boltzmann weight,

\be
w_\beta(x_f)=\sum_R d_\R \chi_\R(x_f) e^{-\beta C_R},
\label{weight}
\ee
to each 2-dimensional face, $f$. The argument is a holonomy along the
face, \ie the ordered product of
gauge variables along a boundary, $\partial f$, of the face $f$:

\be
x_f=\prod_{k\in \partial f} g_k
\label{arg}
\ee
In \eq{arg}, every factor is taken respecting an orientation of links
and faces.
The change of the orientation corresponds to the conjugation $g_k\to
g_k^+$ (or $x_f\to x_f^+$).

By a lattice I mean a cell (polyhedral)
decomposition of a \mbox{$d$-dimensional} manifold such that any
cell can enter in a boundary of another one only once, and every two
cells can border upon each other along only one less dimensional
cell. Simplicial complexes and their duals
obey this restriction by definition. In eq. (\ref{weight}),
$\sum_R$ is the sum over all irreps of the gauge group $G$;
$\chi_\R(x_f)$ is the character of an irrep $R$;
$d_R=\chi_\R(I)$ is its dimension; $C_R$ is a
second Casimir and $\beta$ is a number.
The construction makes sense for compact groups when
unitary finite dimensional irreps span the regular
representation.
Therefore, $R$ is always a discreet index. The choice (\ref{weight})
provides that
$w_\beta(x_f)$ becomes the group $\delta$-function
when $\beta\to 0$:

\be
w_0(x_f)=\delta(x_f,I)
\label{w0}
\ee

The partition function is defined as the integral over all field
configurations:

\be
Z_\beta=\int_G\prod_{\ell} dg_{\ell}\:
\prod_f w_\beta(\prod_{k\in\partial f}g_k)
\label{Z}
\ee
where $dg$ is the Haar measure on the group $G$.

\bigskip

Now, we would like to make the gauge group quantum\footnote{In this
context, the word ``quantum'' may be misleading, but it has already
become standard having actually supplanted the term ``q-deformed''.}. The
simplest example of quantum group is $GL_q(2)$ elements of which can be
defined as

\be
g=\left(\ba{cc} a&b\\ c&d \ea \right)
\label{qmatrix}
\ee
where
\be\ba{lll}
ab=qba & bd=qdb & bc=cb \\
ac=qca & cd=qdc & ad-da=(q-q^{-1})bc
\ea\label{comrel}
\ee
The matrices can be multiplied. If elements of both $g_1$ and $g_2$ obey
\eq{comrel} and are mutually commutative,
the elements of the product obey (\ref{comrel}) as well \cite{Qgroups}.
Therefore, matrices on different links of a lattice have to commute with
one another in the tensor product (as well as with  matrices perfoming gauge
transformations).

The determinant

\be
{\rm Det}_q\ g = ad-qbc
\ee
is central, therefore, one can put it equal to 1. In this way, one
arrives at $SL_q(2)$, which has two real forms: $SU_q(2)$, for real $q$,
and $SL_q(2,R)$, for $|q|=1$ \cite{Qgroups}.
\index{Quantum groups! $SL_q(2)}

The relations (\ref{comrel}) imply the existence of the $R$-matrix

\be
R=\left(\ba{cccc}
q&0&0&0\\ 0&1&0&0\\ 0&q-q^{-1}&1&0\\ 0&0&0&q
\ea\right)\ee
and the $RTT=TTR$ equation

\be
Rg_1\otimes g_2=g_2\otimes g_1R
\label{RTT=TTR}
\ee
$R$ itself obeys the Yang-Baxter equation
\index{Yang-Baxter equation}

\be
R_{12}R_{13}R_{23}=R_{23}R_{13}R_{12}
\label{YB}
\ee
Indices show at which positions in the tensor cube $V\otimes V\otimes V$
acts the $R$-matrix.\index{R-matrix}

For classical gauge groups the self-consistency of the model follows
from the
Peter-Weyl theorem stating that the algebra of regular functions on a
compact group is isomorphic to the algebra of matrix elements of finite
dimensional representations. The quantum version of this theorem was
proven for real $q$'s in Refs.~\cite{Woron}.
In this case there is
the one-to-one correspondence between representations of $SU_q(N)$ and
$SU(N)$, and the notion of the matrix element is naturally generalized.

Therefore, by the space of functions, one may mean a vector space spanned by
matrix elements of irreps, $T_{j,\alpha\beta}(g)$.
\eq{qmatrix} can be regarded as the
fundamental representation. Matrix elements always obey the
$RTT=TTR$ equation (\ref{RTT=TTR}) and, by definition,

\be
T_{j,\alpha\beta}(gh)=\sum_{\gamma=-j}^j
T_{j,\alpha\gamma}(g)T_{j,\gamma\beta}(h)
\ee

The next ingredient is the integral, whose existence is postulated.
It is defined simply as

\be
\int dg\ T_j(g)=\delta_{j,0}
\label{integral}
\ee
\ie whatever is integrated the answer is always zero except for the
trivial representation, which is just a constant.

If one has a product of functions, one can always re-expand products of
matrix elements by using Clebsch-Gordan coefficients:

\be
T_{j_1,\alpha\beta}(g)T_{j_2,\gamma\delta}(g)=
\sum_{j_3=|j_1-j_2|}^{j_1+j_2} \sum_{\sigma,\eps=-j_3}^{j_3}
\langle j_1\alpha,\; j_2\gamma\vert j_3\sigma\rangle T_{j_3,\sigma\eps}
(g)\langle j_3\eps\vert j_1\beta,\; j_2\delta\rangle
\label{decomp}
\ee
Applying this equation successively one can, in principle, reduce
an arbitrary  integral to the basic one (\ref{integral}).

The last ingredient is the character entering the definition of the
weight (\ref{weight}). It should be said that this notion is missing
for $SL_q(2)$. If one naively defines it as the quantum trace of
a matrix element,

\be
\chi_j(g) \stackrel{?}{=}{\rm Tr}_q T_j(g)=\sum_{\alpha=-j}^j
q^{\alpha} T_{j,\alpha\alpha}(g)
\ee
then one finds that $[\chi_j(gh),\chi_i(gf)]\neq0$.
It seems to be impossible to $q$-deform the partition function (\ref{Z})
in a self-consistent way simply starting with this definition!
It is a manifestation of the fact that qQCD$_D$ does not exist at
arbitrary $D$.

On the other hand,  the quantum dimension is equal to
the quantum trace of the identity operator:

\be
[2j+1]\equiv \sum_{\alpha=-j}^j q^{\alpha} =
\frac{q^{j+\frac12}-q^{-j-\frac12}}{q^{\frac12}-q^{-\frac12}}
\ee

However, to define the partition function (\ref{Z}) in the quantum case,
one does not actually need the notion of the character!

The profound correspondence between quantum groups and links of
knots\footnote{I use the term ``link'' to denote 1-dimensional simplexes
as well hoping it should not lead to misunderstanding.} suggests that the
most adequate way to define qQCD$_3$ would be to connect all involved
notions with certain geometric objects. After a projection onto a
plane, the partition function can be given a meaning by putting into
correspondence quantum-group quantities to all geometrical elements. If
one takes another plane, quantum-group symmetries should provide the
independence of the construction from a way of projection. I shall
follow closely \rf{ReshTur}. The basic notion is the tangle, which is
defined as follows. One takes a spherical ball inside which there are a
number of oriented loops and segments whose ends lie on the boundary of
the ball. They are all colored with $SL_q(2)$ representations. One puts
into correspondence to every tangle an operator $O$
acting in the tensor
product of representation spaces $V_{j_1}\otimes\ldots\otimes V_{j_n}$,
(if there are $n$ segments colored $j_1,\ldots,j_n$; their
orientations show the direction of the action of $O$):
\newcounter{cnt}
\setlength{\unitlength}{1mm}
\be
\raisebox{-2cm}{\bp(30,40)(0,-20)
\thicklines
\put(5,-5){\framebox(20,10){O}}
\multiput(7,5)(4,0){5}{\vector(0,1){3}\line(0,1){5}}
\multiput(7,-10)(4,0){5}{\vector(0,1){3}\line(0,1){5}}
\put(5,-12){$\scr j_1\beta_1$}
\put(21,-12){$\scr j_n\beta_n$}
\put(5,12){$\scr j_1\alpha_1$}
\put(21,12){$\scr j_n\alpha_n$}
\ep}
\equiv O_{j_1,\alpha_1\beta_1;\ldots;j_n,\alpha_n\beta_n}
\ee
For example, if there is only one segment and no loops, one finds the
$\delta$-function:

\be
\raisebox{-0.8cm}{\bp(10,20)
\thicklines
\put(5,0){\shortstack{$\alpha$\\ \rule{0cm}{1cm}
\\ \rule{0cm}{0.4cm}$\beta$}}
\put(7,4){\vector(0,1){6}\line(0,1){11}}
\ep}
\equiv \delta_{\alpha,\beta}
\ee
The $R$-matrix distinguishes between under- and over-crossings:

\be
\raisebox{-1cm}{\bp(20,20)
\thicklines
\put(10,10){\line(1,1){7}}
\put(10,10){\line(-1,-1){7}}
\put(11,9){\line(1,-1){6}}
\put(9,11){\line(-1,1){6}}
\ep}
\equiv R=\sum_i a_i\otimes b_i
\ee

\be
\raisebox{-1cm}{\bp(20,20)
\thicklines
\put(10,10){\line(1,-1){7}}
\put(10,10){\line(-1,1){7}}
\put(11,11){\line(1,1){6}}
\put(9,9){\line(-1,-1){6}}
\ep}
=\hspace{0.5cm}
\raisebox{-1cm}{\bp(40,20)
\thicklines
\put(5,0){\line(-1,1){7}}
\put(5,0){\line(1,1){10}}
\put(15,10){\line(1,1){10}}
\put(15,10){\line(-1,-1){10}}
\put(16,9){\line(1,-1){8}}
\put(14,11){\line(-1,1){8}}
\put(25,20){\line(1,-1){7}}
\put(25,20){\line(-1,-1){10}}
\ep}
\equiv R^{-1}=\sum_i b_i\otimes s(a_i)
\ee
where $s$ is the antipod in the ${\cal U}_q(sl(2))$ Hopf algebra.
The Clebsch-Gordan coefficients are represented as the 3-valent vertices

\be
\raisebox{-1cm}{\bp(25,30)
\thicklines
\put(10,10){\line(1,-1){7}}
\put(10,10){\line(-1,-1){7}}
\put(10,10){\line(0,1){10}}
\put(10,22){$j_3\gamma$}
\put(0,0){$j_1\alpha$}\put(17,0){$j_2\beta$}
\ep}
\equiv \langle  j_3\gamma | j_1\alpha, j_2\beta \rangle
\ee

\be
\raisebox{-1cm}{\bp(25,30)
\thicklines
\put(10,10){\line(1,1){7}}
\put(10,10){\line(-1,1){7}}
\put(10,10){\line(0,-1){7}}
\put(7,0){$j_3\gamma$}
\put(0,20){$j_1\alpha$}\put(18,20){$j_2\beta$}
\ep}
\equiv \langle j_1\alpha, j_2\beta | j_3\gamma\rangle
\ee

Matrix elements can be drawn as

\be
\raisebox{-1.4cm}{\bp(10,30)
\thicklines
\put(5,0){\shortstack{$\alpha$\\ \rule{0cm}{2cm}
\\ \rule{0cm}{0.4cm}$j,\beta$}}
\put(7,4){\vector(0,1){4}\line(0,1){8}}
\put(7,18){\vector(0,1){4}\line(0,1){8}}
\put(4,12){\framebox(6,6){$g$}}
\ep}
\equiv T_{j,\alpha\beta}(g)
\ee

The Yang-Baxter and $RTT=TTR$ equations take the familiar graphical
forms

\be
\raisebox{-1cm}{\bp(20,20)
\thicklines
\put(5,5){\line(1,0){15}}
\put(5,5){\line(-1,0){5}}
\put(6,6){\line(1,1){10}}
\put(4,4){\line(-1,-1){4}}
\put(14.5,5.5){\line(-1,1){4}}
\put(16,4){\line(1,-1){4}}
\put(9,11){\line(-1,1){4.5}}
\ep}
=
\raisebox{-1cm}{\bp(20,20)
\thicklines
\put(5,15){\line(1,0){15}}
\put(5,15){\line(-1,0){5}}
\put(4,16){\line(-1,1){4}}
\put(5.5,14.5){\line(1,-1){4}}
\put(16,16){\line(1,1){4}}
\put(14,14){\line(-1,-1){9}}
\put(11,9){\line(1,-1){4}}
\ep}
\ee
and

\be
\raisebox{-1.5cm}{\bp(15,30)
\thicklines
\put(3,4){\line(0,1){8}}
\put(3,18){\line(0,1){3}}
\put(3,21){\line(1,1){8}}
\put(0,12){\framebox(6,6){$g_1$}}
\put(11,4){\line(0,1){8}}
\put(11,18){\line(0,1){3}}
\put(11,20){\line(-1,1){4}}
\put(6,25){\line(-1,1){4}}
\put(8,12){\framebox(6,6){$g_2$}}
\ep}
=
\raisebox{-1.5cm}{\bp(15,30)
\thicklines
\put(3,9){\line(0,1){3}}
\put(2.5,9.5){\line(1,-1){4}}
\put(7.5,4.5){\line(1,-1){4}}
\put(3,18){\line(0,1){8}}
\put(0,12){\framebox(6,6){$g_2$}}
\put(11,9){\line(0,1){3}}
\put(11,9){\line(-1,-1){8}}
\put(11,18){\line(0,1){8}}
\put(8,12){\framebox(6,6){$g_1$}}
\ep}
\ee

One needs also the quantum trace of an operator, which is equivalent to
the closure of a tangle

\be
\raisebox{-1.8cm}{\bp(40,40)(0,-20)
\thicklines
\put(15,-5){\framebox(20,10){O}}
\multiput(17,5)(4,0){5}{\line(0,1){3}}
\multiput(17,-8)(4,0){5}{\line(0,1){3}}
\put(13,8){\oval(8,4)[t]}
\put(13,-8){\oval(8,4)[b]}
\put(9,-8){\line(0,1){16}}
\put(17,8){\oval(32,10)[t]}
\put(17,-8){\oval(32,10)[b]}
\put(1,-8){\line(0,1){17}}
\put(20,10){\ldots\ldots}
\put(20,-10){\ldots\ldots}
\put(2,0){\ldots}
\ep}
\equiv \sum_{\alpha_1=-j_1}^{j_1}\ldots
\sum_{\alpha_n=-j_n}^{j_n} \prod_{i=1}^{n} q^{\alpha_i}\;
O_{j_1,\alpha_1\alpha_1;\ldots;j_n,\alpha_n\alpha_n}
\label{closure}
\ee

To each link of the lattice, one puts into correspondence an integral of
a product of matrix elements, the number of which is equal to the number
of faces incident to the link. One can associate a tangle with every
such integral. It means a cell decomposition of the manifold. The
partition function can be constructed by connecting these tangles
together or, equivalently, by gluing up the 3-cells. There appears an
index loop going along a boundary of every face. In three dimensions,
there is a natural cyclic order of faces sharing the same link. The
index loops have to be ordered according to it. After that the partition
function can be unambiguously defined.

If $q=e^{i\frac{2\pi}{k+2}}$, one has to restrict all indices to the
fusion ring: \mbox{$j=0,\frac12,1,\ldots,\frac{k}2$}. In this case the
following tangle can serve as the definition of the matrix element

\be
T_{j,\alpha\beta}(g)\equiv\
\raisebox{-0.9cm}{\bp(10,20)
\thicklines
\put(3,0){\shortstack{$\scr j,\alpha$\\ \rule{0cm}{1.2cm}
\\ \rule{0cm}{0.4cm}$\scr j,\beta$}}
\put(5,5){\line(0,1){3}}
\put(5,10){\line(0,1){6}}
\put(0,9){\line(1,0){10}}
\put(0,11){\line(1,0){4}}
\put(6,11){\line(1,0){4}}
\ep}
\ee

The tensor product of matrix elements looks as

\be
T_{j_1,\alpha_1\beta_1}(g)T_{j_2,\alpha_2\beta_2}(g)\equiv\
\raisebox{-0.9cm}{\bp(20,20)
\thicklines
\put(3,0){\shortstack{$\scr j_1,\alpha_1$\\ \rule{0cm}{1.2cm}
\\ \rule{0cm}{0.4cm}$\scr j_1,\beta_1$}}
\put(12,0){\shortstack{$\scr j_2,\alpha_2$\\ \rule{0cm}{1.2cm}
\\ \rule{0cm}{0.4cm}$\scr j_2,\beta_2$}}
\put(5,3){\line(0,1){5}}
\put(5,10){\line(0,1){6}}
\put(15,3){\line(0,1){5}}
\put(15,10){\line(0,1){6}}
\put(0,9){\line(1,0){20}}
\put(0,11){\line(1,0){4}}
\put(6,11){\line(1,0){8}}
\put(16,11){\line(1,0){4}}
\ep}
\ee
and the integral takes the form of the finite sum

\[
\int dg\ T_{j,\alpha\beta}(g)\equiv d_0\sum_{i=0}^{k/2} d_i\
\raisebox{-0.9cm}{\bp(15,20)
\thicklines
\put(3,0){\shortstack{$\scr j,\alpha$\\ \rule{0cm}{1.2cm}
\\ \rule{0cm}{0.4cm}$\scr j,\beta$}}
\put(5.5,3){\line(0,1){4}}
\put(5.5,9){\line(0,1){8}}
\put(3,8){\line(1,0){5}}
\put(3,12){\line(1,0){2}}
\put(6,12){\line(1,0){2}}
\put(3,10){\oval(4,4)[l]}
\put(8,10){\oval(4,4)[r]}
\put(11,11){$\scr i$}
\ep}=\hspace{4cm}
\]\be
\frac{2\sin\frac{\pi}{k+2}}{k+2}\sum_{i=0}^{k/2}
\sin\frac{\pi(2i+1)}{k+2}\;
\frac{\sin\frac{\pi(2i+1)(2j+1)}{k+2}}{\sin\frac{\pi(2j+1)}{k+2}}\;
\delta_{\alpha,\beta}=\delta_{j,0}\delta_{\alpha,0}\delta_{\beta,0}
\label{qint}
\ee
where $d_j$ is the quantum dimension conveniently normalized:

\be
d_j=\sqrt{\frac2{k+2}}\sin\frac{\pi(2j+1)}{k+2}
\ee

To prove \eq{qint}, I used results of Reshetikhin and Turaev
\cite{ReshTur}. My claim
is that it can be regarded as the definition of the integral on the
fusion ring of
$SL_q(2)$, $q=e^{i\frac{2\pi}{k+2}}$.

In addition to the fusion ring irreps $\{V_j\}$,
$j=0,\frac12,\ldots,\frac{k}2$; ${\cal U}_q(sl(2))$ has a number of
representations having the vanishing quantum dimension \cite{sl2reps}:
$\{I_p\}$, \mbox{$p=-\frac12,0,\frac12,1,\ldots, \frac{k+1}2$}.
Representations $I_p$ for $0\leq p\leq\frac{k+1}2$ although
not irreducible are indecomposable.

The tensor product of two irreps
from the fusion ring has the following decomposition

\be
V_i\otimes V_j = \Big(\bigoplus_{m=|i-j|}^{{\rm min}(i+j,k-i-j)}
V_m\Big)\oplus
\Big(\bigoplus_{\mbox{}^{-\frac12\leq p \leq i+j-\frac{k+2}2}
_{\ p=(i+j)\, \bmod\; 1}}
I_p\Big)
\ee
The set of representations $\{I_p\}$ forms an ideal

\be
\{V_j\}\otimes\{I_p\}\subset\{I_p\} \hspace{1cm}
\{I_p\}\otimes\{I_p\}\subset\{I_p\}
\ee

As was proven by Reshetikhin and Turaev \cite{ReshTur},
the closure of any tangle
vanishes if at least one representation of this type appears in it. More
precisely, they have shown that any ${\cal U}_q(sl(2))$-linear operator
acting in $\{I_p\}$ has the vanishing quantum trace. Obviously, it holds
for operators obtained by cutting an internal line in an arbitrary
closed tangle. As it takes place for any line, colors
from the set $\{I_p\}$ never appear. Therefore, when all index
loops are closed, these representations can be simply ignored.
Thus, one has the following orthogonality property

\[
\int dg\ T_{j_1,\alpha_1\beta_1}(g)T_{j_2,\alpha_2\beta_2}(g)\equiv\
d_0\sum_{i=0}^{k/2} d_i\
\raisebox{-0.9cm}{\bp(20,20)
\thicklines
\put(3,0){\shortstack{$\scr j_1,\alpha_1$\\ \rule{0cm}{1.2cm}
\\ \rule{0cm}{0.4cm}$\scr j_1,\beta_1$}}
\put(12,0){\shortstack{$\scr j_2,\alpha_2$\\ \rule{0cm}{1.2cm}
\\ \rule{0cm}{0.4cm}$\scr j_2,\beta_2$}}
\put(5.5,3){\line(0,1){4}}
\put(5.5,9){\line(0,1){8}}
\put(15,3){\line(0,1){4}}
\put(15,9){\line(0,1){8}}
\put(2,8){\line(1,0){16}}
\put(2,12){\line(1,0){3}}
\put(6,12){\line(1,0){8}}
\put(16,12){\line(1,0){2}}
\put(2,10){\oval(4,4)[l]}
\put(18,10){\oval(4,4)[r]}
\ep}\ =
\]\be
\sum_{m=|j_1-j_2|}^{{\rm min}(j_1+j_2,k-j_1-j_2)}
d_0\sum_{i=0}^{k/2} d_i\
\raisebox{-0.9cm}{\bp(20,20)
\thicklines
\put(3,0){\shortstack{$\scr j_1,\alpha_1$\\ \rule{0cm}{1.2cm}
\\ \rule{0cm}{0.4cm}$\scr j_1,\beta_1$}}
\put(12,0){\shortstack{$\scr j_2,\alpha_2$\\ \rule{0cm}{1.2cm}
\\ \rule{0cm}{0.4cm}$\scr j_2,\beta_2$}}
\put(10,6){\line(0,1){2}}
\put(10,9){\line(0,1){5}}
\put(10,14){\line(1,1){4}}
\put(10,14){\line(-1,1){4}}
\put(10,6){\line(-1,-1){4}}
\put(10,6){\line(1,-1){4}}
\put(4,8.5){\line(1,0){12}}
\put(4,12.5){\line(1,0){5}}
\put(11,12.5){\line(1,0){5}}
\put(4,10.5){\oval(4,4)[l]}
\put(16,10.5){\oval(4,4)[r]}
\ep}=
\frac{d_0\delta_{j_1,j_2}}{d_{j_1}}
\raisebox{-0.9cm}{\bp(20,20)
\thicklines
\put(2,2){\shortstack{$\scr \alpha_1$\\ \rule{0cm}{1cm}
\\ \rule{0cm}{0.4cm}$\scr \beta_1$}}
\put(12,2){\shortstack{$\scr \alpha_2$\\ \rule{0cm}{1cm}
\\ \rule{0cm}{0.4cm}$\scr \beta_2$}}
\put(8,5){\oval(10,6)[t]}
\put(8,17){\oval(10,6)[b]}
\ep}
\label{orthog}
\ee
which allows for a Fourier decomposition of an arbitrary function
spanned by matrix elements from the fusion ring. It gives an
analog of the Peter-Weyl theorem. A reader must realize
that Eqs.~(\ref{qint}) and (\ref{orthog}) do not hold for
$\{I_p\}$ representations. For self-consistency, all greek indices have to be
summed over to form a link of 3-valent graphs and loops. {\em All equalities
between tangles have to be understood as taking place after closing
with an arbitrary tangle.}

So, we arrive at the following definition of qQCD$_3$ partition
function on a 3-manifold ${\cal M}$ \cite{B1}:

\be
Z_{\beta}({\cal M})= d_0^{\scs N_3+N_0-2}
\sum_{\{j_f\}}\prod_{f=1}^{N_2} [d_{j_f}c_{j_f}(\beta)]
\sum_{\{j_{\ell}\}}\prod_{\ell=1}^{N_1} d_{j_{\ell}}\;
J_{\{j_{\ell}\},\{j_f\}}({\cal L})
\label{Z=J}
\ee
where $J_{\{j_{\ell}\},\{j_f\}}({\cal L})$ is the Jones polynomial for a
link ${\cal L}$ defined by a cell decomposition of ${\cal M}$. This
link consists \index{Jones polynomial}
of $N_1+N_2$ unframed loops colored with sets of
representations $\{j_{\ell}\}$ and $\{j_f\}$. Loops from the first set,
$\{j_{\ell}\}$, go around 1-cells (links) pinching bunches of loops from
the second set, $\{j_f\}$,  which go along boundaries of 2-cells
(faces); $c_j(\beta)$'s are numbers (weights).
If all $c_j\equiv 1$, $Z_0$ is a
topological invariant. In this case, the partition
function (\ref{Z=J}) is obviously self-dual with respect to the
Poincar\'{e} duality of complexes.

\eq{Z=J} is just a particular implementation of the general
Reshetikhin-Turaev construction \cite{ReshTur}. However, the link $\cal
L$ here is not related to a surgery representation of the manifold.

\index{Turaev-Viro invariant}
In order to establish a connection with the Turaev-Viro invariant, let us
consider lattices dual to simplicial complexes. Their 1-skeletons are
4-valent graphs and exactly 3 faces are incident to each link giving
the integral of 3 matrix elements for every triangle
in a simplicial complex:

\be
\int dg\; T_{j_1,\alpha_1\beta_1}(g)
T_{j_2,\alpha_2\beta_2}(g) T_{j_3,\alpha_3\beta_3}(g)
= \frac{d_0}{d_{j_3}}
\raisebox{-1cm}{\bp(25,20)
\thicklines
\put(2,2){\shortstack{$\scr j_1,\alpha_1$\\ \rule{0cm}{1cm}
\\ \rule{0cm}{0.4cm}$\scr j_1,\beta_1$}}
\put(10,2){\shortstack{$\scr j_2,\alpha_2$\\ \rule{0cm}{1cm}
\\ \rule{0cm}{0.4cm}$\scr j_2,\beta_2$}}
\put(18,2){\shortstack{$\scr j_3,\alpha_3$\\ \rule{0cm}{1cm}
\\ \rule{0cm}{0.4cm}$\scr j_3,\beta_3$}}
\put(9,4){\oval(8,6)[t]}
\put(9,18){\oval(8,6)[b]}
\put(15,7){\oval(12,4)[lt]}
\put(15,15){\oval(12,4)[lb]}
\put(15,5){\oval(12,8)[rt]}
\put(15,17){\oval(12,8)[rb]}
\ep}
\label{3j}
\ee
The right hand side of eq. (\ref{3j}) is the product of two 3-$j$
symbols. Summing over lower indices one gets a Racah-Wigner 6-$j$
symbol

\be
\left\{\ba{ccc}j_1&j_2&j_3\\j_4&j_5&j_6\ea\right\} =
\frac{d_0^2}{d_{j_6}\sqrt{d_{j_2}d_{j_5}}}
\raisebox{-2cm}{\bp(40,40)(-5,-5)
\thicklines
\put(5,15){\oval(10,10)[b]}
\put(12.5,10){\oval(15,10)[b]}
\put(17.75,5){\oval(14.5,10)[b]}
\put(7.5,20){\oval(15,10)[t]}
\put(15,15){\oval(10,10)[t]}
\put(16.25,25){\oval(17.5,10)[t]}
\put(0,15){\line(0,1){5}}
\put(20,10){\line(0,1){5}}
\put(25,5){\line(0,1){20}}
\put(1,26){$\scr j_1$}
\put(8,18){$\scr j_3$}
\put(15,24){$\scr j_2$}
\put(16,8){$\scr j_4$}
\put(5,3){$\scr j_5$}
\put(27,15){$\scr j_6$}
\ep}
\label{6j}
\ee
inside each
tetrahedron of a simplicial complex. Representation indices, $j_f$, are
attached to its 1-simplexes, $f$, ({\em i.e.} faces of the dual lattice).
The partition function $Z_{0}$ can be written then in the Turaev-Viro
form

\be
Z_0 = d_0^{\scs N_1-N_2-2} \sum_{\{j_f\}} \prod_{f=1}^{N_2} d_{j_f}\,
\prod_{t=1}^{N_0}
\left\{\ba{ccc}j_{t_1}&j_{t_2}&j_{t_3}\\j_{t_4}&j_{t_5}&j_{t_6}\ea\right\}
\label{TuV}
\ee
where the indices $t_1,\ldots,t_6$ denote six edges of a $t$'th
tetrahedron.

To prove that $Z_0$ is indeed a topological invariant it is
sufficient to show that it is unchanged under the moves shown in
Figure~1. However, the link representation (\ref{Z=J}) is more
convenient in this respect. By using the analog of the group measure
invariance

\be
\int dg\; f(gh)\equiv\
d_0\sum_{i=0}^{k/2} d_i\
\raisebox{-1.5cm}{\bp(20,30)
\thicklines
\put(5.5,3){\line(0,1){4}}
\put(5.5,9){\line(0,1){8}}
\put(15.5,3){\line(0,1){4}}
\put(15.5,9){\line(0,1){8}}
\put(2,8){\line(1,0){16}}
\put(2,12){\line(1,0){3}}
\put(6,12){\line(1,0){9}}
\put(16,12){\line(1,0){2}}
\put(2,10){\oval(4,4)[l]}
\put(18,10){\oval(4,4)[r]}
\put(2,18){\line(1,0){16}}
\put(5.5,19){\line(0,1){6}}
\put(15.5,19){\line(0,1){6}}
\put(2,21){\line(1,0){3}}
\put(6,21){\line(1,0){9}}
\put(16,21){\line(1,0){2}}
\put(7.5,10){\ldots}
\put(7.5,19.5){\ldots}
\ep}\ =
d_0\sum_{i=0}^{k/2} d_i\
\raisebox{-1.5cm}{\bp(20,30)
\thicklines
\put(5.5,3){\line(0,1){4}}
\put(5.5,9){\line(0,1){7}}
\put(15.5,3){\line(0,1){4}}
\put(15.5,9){\line(0,1){7}}
\put(2,8){\line(1,0){16}}
\put(2,12){\line(1,0){3}}
\put(6,12){\line(1,0){9}}
\put(16,12){\line(1,0){2}}
\put(2,10){\oval(4,4)[l]}
\put(18,10){\oval(4,4)[r]}
\put(2,17){\line(1,0){16}}
\put(5.5,22){\line(0,1){3}}
\put(15.5,22){\line(0,1){3}}
\put(5.5,20){\line(0,-1){2}}
\put(15.5,20){\line(0,-1){2}}
\put(2,21){\line(1,0){16}}
\put(8,10){\ldots}
\put(8,19){\ldots}
\ep}\ =
\int dg\; f(g)
\ee
one can reduce the number of loops. The corresponding operations
have a nice interpretation as
topology preserving transformations of complexes \cite{B1}.

For lattices, as they have been defined above, all loops are
unframed. However, transforming complexes, one can obtain non-trivial
framings and, in real calculations, has to follow them carefully.
Practically, it is convenient to use the
ribbon graph representation \cite{ReshTur}. The framing of a ribbon loop
is defined as a linking number of its edges. It is fixed by the
condition that one of two sides of the ribbon is always turned toward the
inside of a 2-cell which it encircles (or toward a 1-cell which it
wraps).

The invariant is multiplicative with respect to the connected sum of
complexes
\be
Z_0(C_1\# C_2)=Z_0(C_1)Z_0(C_2)
\ee
because for the sphere
\be
Z_0(S^3)=1
\ee

Every oriented complex can be
transformed into the canonical form, when there are single 0- and
3-dimensional cells and the equal number, $\nu$, of 1- and 2-cells:

\be
C=\sigma^0\cup
(\bigcup_{i=1}^{\nu} \sigma^1_i)\cup
(\bigcup_{j=1}^{\nu}\sigma^2_j)\cup
\sigma^3
\ee
One can put into correspondence with each 1-cell, $\sigma^1_i$, a
generator of the fundamental group $\gamma_i\in\pi_1(C)$. Each 2-cell,
$\sigma^2_j$, gives a
defining relation for $\pi_1(C)$:

\be
\Gamma_j=\prod_{\sigma^1_k\in\partial\sigma^2_j}\gamma_k=I
\label{defrel}
\ee

If the gauge group is a classical finite group $G$, the partition function
$Z_0$ is well defined (after substituting the sum
$\sum_{g\in G}$ for $\int dg$):

\be
Z_0^{\scs(G)}=\sum_{\{g_i\}} \prod_{j=1}^{\nu}
\delta(\prod_{\sigma^1_k\in\partial\sigma^2_j}g_k,I)
\label{Z0(G)=}
\ee
This expression equals the number of representations of the fundamental group
by elements of the gauge one: $\pi_1(C)\to G$. Hence, it is an integer.
In the quantum case, a similar interpretation exists. One have to
consider an action of the fundamental group on the universal covering of
a complex. It acts permuting cells of the covering, which can be
regarded as a $\pi(C)$-module.
The invariant can be said to be the ``quantum analog'' of \eq{Z0(G)=},
where the $\pi(C)$-action on the universal covering is represented
by elements of a quantum gauge group. It is a real number.

As was proven by Turaev \cite{Tur},
the Turaev-Viro invariant is equal to the
Reshetikhin-Turaev-Witten one modulo squared:
$Z_0({\cal M})=|I({\cal M})|^2$. Kohno \cite{Kohno} has shown
that it is bounded from above as

\be
Z_0({\cal M})\leq \Big(\frac1{d_0}\Big)^{2h}
\ee
where $h$ is a Heegaar genus of ${\cal M}$, \ie the minimum genus of
handlebodies appearing in Heegaar splittings of $\cal M$.

\section{Generating function for simplicial complexes.}

In Refs.~\cite{B2,B1} the zero-dimensional field model
generating all possible
simplicial complexes weighted with the partition function (\ref{TuV})
was suggested. Let
$\phi(x,y,z)$ be a function on $G\otimes G\otimes G$ invariant under
right shifts

\be
\phi(x,y,z)=\phi(xu,yu,zu)\hspace{1cm}\forall x,y,z,u\in G
\label{shiftinvar}
\ee
and symmetric under even permutations. Odd ones are equivalent to the
complex conjugation:

\be
\phi(x,y,z)=\phi(y,z,x)=\phi(z,x,y)=\overline{\phi}(y,x,z)
\ee
It can be represented in terms of matrix elements as

\[
\phi(x,y,z)=\sum_{ \mbox{}^{j_1j_2j_3}_{\, \{a_i,b_i\}}}
\sqrt{d_{j_1}d_{j_2}}
\varphi^{j_1j_2j_3}_{a_1a_2a_3}
T_{j_1,a_1b_1}(x)T_{j_2,a_2b_2}(y)T_{j_3,a_3b_3}(z)
\left(\ba{ccc} j_1&j_2&j_3\\ b_1&b_2&b_3\ea\right)
\]\be
=\sum_{\{j_i,a_i,b_i\}}
\sqrt{\frac{d_{j_1}d_{j_2}}{d_{j_3}}}
\raisebox{-1cm}{\bp(25,25)
\put(2,10){\framebox(3,3){$\scr x$}}
\put(10,10){\framebox(3,3){$\scr y$}}
\put(18,10){\framebox(3,3){$\scr z$}}
\put(6,16){\framebox(3,3){$\scr \varphi$}}
\thicklines
\put(7.5,10){\oval(8,6)[b]}
\put(13.5,7){\oval(12,4)[lb]}
\put(13.5,10){\oval(12,10)[rb]}
\put(19.5,13){\line(0,1){6}}
\put(13.5,19){\oval(12,7)[lt]}
\put(13.5,18){\oval(12,9)[rt]}
\put(5.5,13){\oval(4,7)[lt]}
\put(9.5,13){\oval(4,7)[rt]}
\ep}
\ee
where $\left(\ba{ccc} j_1&j_2&j_3\\ b_1&b_2&b_3\ea\right)$ is the 3-$j$
symbol; $\varphi^{j_1j_2j_3}_{a_1a_2a_3}=
\overline{\varphi}^{j_2j_1j_3}_{a_2a_1a_3}$ and symmetric under cyclic
permutations.
This equation is a general Fourier decomposition of a function
obeying (\ref{shiftinvar}).

The partition function is defined as the integral

\be
P=\int {\cal D}\phi\; e^{-S}
\label{parfun2}
\ee
where the action is taken in the form

\[
S=\frac{1}{2}\int dxdydz\; |\phi(x,y,z)|^2-\hspace{4cm}
\]\be
\frac{\lambda}{12}
\int dxdydzdudvdw\; \phi(x,y,z)\phi(x,u,v)\phi(y,v,w)\phi(z,w,u)
\label{action}
\ee
The first term in eq. (\ref{action}) can be imagined as two glued
triangles and the second, as four triangles forming a tetrahedron. It is
not surprising that, after the Fourier transformation, one finds a 6-$j$
symbol associated with it:

\be
S=\frac12\sum_{ j_1j_2j_3}\frac1{d_{j_3}}
\raisebox{-1cm}{\bp(20,20)
\put(6,16){\framebox(3,3){$\scr \varphi$}}
\put(6,6){\framebox(3,3){$\scr \varphi$}}
\thicklines
\put(5.5,12.5){\oval(4,8)[l]}
\put(9.5,12.5){\oval(4,8)[r]}
\put(12.5,19){\oval(10,7)[t]}
\put(12.5,6){\oval(10,7)[b]}
\put(17.5,6){\line(0,1){13}}
\ep}
-\frac{\lambda}{12}
\sum_{j_1\ldots j_6}\frac{1}{d_{j_1}^2d_{j_2}d_{j_3}}
\raisebox{-1cm}{\bp(25,25)
\put(2,10){\framebox(3,3){$\scr \varphi$}}
\put(10,10){\framebox(3,3){$\scr \varphi$}}
\put(18,10){\framebox(3,3){$\scr \varphi$}}
\put(6,16){\framebox(3,3){$\scr \varphi$}}
\thicklines
\put(7.5,10){\oval(5,6)[b]}
\put(15.5,10){\oval(5,6)[b]}
\put(11.25,10){\oval(19,14)[b]}
\put(19.5,13){\line(0,1){6}}
\put(13.5,19){\oval(12,7)[lt]}
\put(13.5,18){\oval(12,9)[rt]}
\put(5.5,13){\oval(4,7)[lt]}
\put(9.5,13){\oval(4,7)[rt]}
\ep}
\raisebox{-1cm}{\bp(25,25)
\thicklines
\put(3.5,10){\oval(3,6)[t]}
\put(11.5,10){\oval(3,6)[t]}
\put(19.5,10){\oval(3,6)[t]}
\put(7.5,10){\oval(5,6)[b]}
\put(15.5,10){\oval(5,6)[b]}
\put(11.25,10){\oval(19,14)[b]}
\put(19.5,13){\line(0,1){3}}
\put(13.5,16){\oval(12,13)[t]}
\put(7.5,13){\oval(7,7)[t]}
\ep}
\label{graphact}
\ee
The measure can be written in terms of Fourier coefficients

\be
{\cal D}\phi =\prod_{\mbox{}^{j_1j_2j_3}_{a_1a_2a_3}}
d\varphi^{j_1j_2j_3}_{a_1a_2a_3}
\label{measure}
\ee
If $q=e^{i\frac{2\pi}{k+2}}$,
the product in eq. (\ref{measure}) runs over irreps from
the fusion ring and, hence, is finite.

Practically, the partition function (\ref{parfun2}) has a meaning
within the perturbation expansion in $\lambda$.
Performing all possible Wick pairings,
one gets in every order in $\lambda$ all oriented simplicial complexes.
For every 1-simplex in a simplicial complex,
one has a loop carrying a representation index.
It gives a corresponding quantum dimension.
A 6-$j$ symbol inside each tetrahedron has already appeared in
eq. (\ref{graphact}). Summing over all representations on links,
one reproduces the Turaev-Viro partition
function for a given simplicial complex.

Therefore, $\log P$ is a generating function of 3D simplicial complexes
weighted with the Turaev-Viro invariant. Of course, $P$ is only formally
defined.  However, this construction gives a framework for the strong
coupling expansion in simplicial gravity, which can be carried out by
iterating the Schwinger-Dyson equation for the partition function
(\ref{parfun2}) \cite{Ooguri2}.

A more down-to-earth model can be obtained by taking classical finite
gauge group.
Repeating all steps, one finds the sum over all simplicial
complexes weighted with the invariant (\ref{Z0(G)=}) times a volume
dependent factor. To make a contact with the discrete action
(\ref{Slatt}),  one has to introduce a fugacity $\mu$
for the number of links as
well. It can be done by adding three indices to $\phi$:

\[
\log P^{(G)}= \int \prod_{\{x_i\in G\}} \prod_{k_i=1}^{\mu}
d\phi^{k_1k_2k_3}_{x_1x_2x_3}
\exp\Big\{-\frac{1}{2}\sum_{\{x_i\in G\}} \sum_{k_i=1}^{\mu}\;
|\phi^{k_1k_2k_3}_{x_1x_2x_3}|^2+
\]\[
\frac{\lambda}{12}\sum_{\{x_i\in G\}} \sum_{k_i=1}^{\mu}
\; \phi^{k_1k_2k_3}_{x_1x_2x_3}\phi^{k_1k_4k_5}_{x_1x_4x_5}
\phi^{k_2k_5k_6}_{x_2x_5x_6}\phi^{k_3k_6k_4}_{x_3x_6x_4}=
\]\be
\sum_{\{C\}} |G|^{N_0-1}\lambda^{N_3}\mu^{N_1} Z_0^{(G)}(C)
\ee
where $|G|=\sum_{G}1$ is the rank of the group.

It can be easily seen that simplicial complexes have non-negative Euler
characters \index{Euler character}

\be
\chi=\sum_{i=1}^{N_0} p_i \geq 0
\label{euler}
\ee
where the sum runs over all vertices. Tetrahedra touching
the $i$'th vertex form a
3D ball; $p_i$ is the genus of its 2D boundary.
By definition, a complex is a manifold {\em iff} $p_i=0\ \forall i$;
\ie the vicinity of every point is a spherical ball.

After the rescaling, $\lambda=|G|\widetilde{\lambda}$,
$\mu=\frac1{|G|}\widetilde{\mu}$, one obtains

\be
|G|\log P^{(G)} = \sum_{\{C\}}
\widetilde{\lambda}^{N_3}\widetilde{\mu}^{N_1}|G|^{\chi} Z_0^{(G)}(C)
\ee
and in the formal limit $|G|\to0$ only manifolds for which $Z_0^{(G)}$
is finite contribute.

Any finite group can be embedded in the permutation group, ${\cal S}_n$,
for sufficiently large $n$; $|G|=n!$ in this case.
It suggests that, at $\widetilde{\lambda}$ and $\widetilde{\mu}$ fixed,
one should take $n$ much bigger than the maximum rank
of the fundamental group for typical complexes and try to continue
analytically to $n!=0$ (But how to do it practically?!). If the
Poincar\'{e} hypothesis is true, only spheres should survive in this
limit. Technically, it could mean a kind of double scaling.

Unfortunately, the model seems to be too complicated to be investigated
analytically.

\section{Conclusion}

My aim in the present paper has been to draw attention to the quite
promising problem of 3D quantum gravity. What one could learn from it
concerns fundamental properties of the quantum vacuum.
Non-renormalizability of gravity, non-boundedness of the
Einstein-Hilbert action, topology changing processes, cosmological
constant problem can be addressed within this simplified (comparing to
4$D$ gravity) framework. Three dimensional geometry and topology possess
a lot of beautiful mathematical structures. Many fundamental and long
standing problems have not yet been solved. It is still a field of
intensive research, which create an exciting atmosphere of a parallel
rise of mathematical results and physical understanding.

\vspace{1cm}

{\Large \bf Acknowledgments}
\bigskip

I would like to express my gratitude to the organizers for the creative
and friendly atmosphere at Cargese during the workshop. At different
stages of work on problems touched in this paper, I have enjoyed the
collaboration and discussions with M.Agistein, A.Alekseev,
J.Ambj\o rn, C.Bachas, C.Itzykson, V.Kazakov, A.N.Kirillov, I.Kostov,
A.Krzywicki, A.A.Migdal, M.Petropoulos and
S.Piunikhin. I appreciate the financial support from the EEC grant
CS1-D430-C and from the Danish National Research Council.


\begin{thebibliography}{99}


\bibitem{David} F.David, {\em ``Simplicial quantum gravity
and random lattices''}, Les Houches lectures, Session LVII (1992).

\bi{Witten}{E.Witten}{Nucl. Phys.}{B311}{1988/89}{46} and {\bf B323}
(1989) 113.

\bibitem{TV} V.G.Turaev and O.Y.Viro, {\sl Topology} {\bf 31} (1992) 865.

\bibitem{Ooguri} H.Ooguri and N.Sasakura, {\sl Mod. Phys. Lett.} {\bf A6}
(1991) 3591;\\
F.Archer and R.M.Williams, {\sl Phys.Lett.} {\bf B273} (1991) 438.

\bibitem{PR} G.Ponzano and T. Regge, {\sl in Spectroscopic and group
theoretical methods in physics}, ed. F.Bloch (North-Holland, Amsterdam,
1968).

\bibitem{Regge} T. Regge, {\sl Nuovo Cimento} 19 (1961) 558.

\bibitem{3dsim} M.E. Agishtein and A.A. Migdal, {\sl Mod. Phys. Lett. }
{\bf A6} (1991) 1863;\\
J.Ambj\o rn and S. Varsted, {\sl Phys. Lett.} {\bf B226} (1991) 258 and
{\sl Nucl. Phys.} {\bf B373} (1992) 557.

\bibitem{3dsim2}D.V. Boulatov and A. Krzywicki, {\sl Mod. Phys. Lett.} {\bf A6}
(1991) 3005;\\
J.Ambj\o rn, D.V. Boulatov, A. Krzywicki and S. Varsted,
{\sl Phys. Lett.} {\bf B276} (1992) 432.

\bi{3dsim3} {J.Ambj\o rn, Z.Burda, J.Jurkiewicz and C.F.Kristjansen}
{Phys. Lett.}{B297}{1992}{253}.

\bi{4dsim}{M.Agishtein and A.A.Migdal}{Mod. Phys. Lett.}{A7}{1992}{1039};\\
J.Ambj\o rn and J.Jurkiewicz, {\sl Phys. Lett.} {\bf B278} (1992) 42.

\bibitem{Nash} See for example, C.Nash, {\it Differential topology and
quantum field theory}, Academic Press; chap.~1 and references therein;

\bibitem{Wilson} K.Wilson, {\sl Phys. Rev.} {\bf D10} (1975) 2445.

\bi{Qgroups}{L.D.Faddeev, N.Reshetikhin and L.Takhtajan}{Leningrad
Math. J}{1}{1990}{193}.

\bibitem{Woron} S.L.Woronowicz, {\sl Commun. Math. Phys.} {\bf 111}
(1987) 613;\\
L.L.Vaksman and Ya.S.Soibelman, {\sl Func. Anal. Appl.} {\bf 22} (1988)
170.

\bibitem{ReshTur} N.Yu.Reshetikhin and V.G.Turaev, {\sl Commun. Math.
Phys.} {\bf 124} (1989) 307 and {\sl Invent. Math.} {\bf 103} (1991) 547.

\bibitem{sl2reps}
P.Roche and D.Arnaucon, {\sl Lett. Math. Phys.} {\bf 17} (1989) 295;\\
V.Pasquier and H.Saleur, {\sl Nucl. Phys.} (1990);\\
G.Keller, {\sl Lett. Math. Phys.} {\bf 21} (1991) 273.

\bibitem{B1} D.V. Boulatov, {\sl Int. J. Mod. Phys.} {\bf A8} (1993) 3139.

\bibitem{Tur} V.G.Turaev, {\sl C.R. Acad. Sci. Paris} {\bf 313} (1991)
395; {\sl J. Diff. Geom.} {\bf 36} (1992) 35.

\bibitem{Kohno} T.Kohno, {\sl Topology} {\bf 31} (1992) 203.

\bibitem{B2} D.V. Boulatov, {\sl  Mod. Phys. Lett.} {\bf A7} (1992) 1629.

\bi{Ooguri2}{H.Ooguri}{Prog. Theor. Phys.}{89}{1993}{1}.

\end{thebibliography}
\end{document}